\documentclass[12pt]{article}
\usepackage{amssymb,amsmath,epsfig,color,mathtools,nccmath}
 \allowdisplaybreaks

\begin{document}
\title{\bf Junction conditions in perfect fluid $f(\mathcal{G},~T)$ gravitational theory}
\author{M. Z. Bhatti \thanks{mzaeem.math@pu.edu.pk}, Z. Yousaf \thanks{zeeshan.math@pu.edu.pk} and M. Yousaf \thanks{myousaf.math@gmail.com}\\
Department of Mathematics, University of the Punjab,\\
Quaid-i-Azam Campus, Lahore-54590, Pakistan.}

\date{}

\maketitle
\begin{abstract}
This manuscript aims to establish the gravitational junction conditions(JCs) for the $f(\mathcal{G},~T)$ gravity. In this gravitational theory, $f$ is an arbitrary function of Gauss-Bonnet invariant $\mathcal{G}$ and the trace of the energy-momentum tensor $T_{\mu\nu}$ i.e., $T$. We start by introducing this gravity theory in its usual geometrical representation and posteriorly obtain a dynamically equivalent scalar-tensor demonstration on which the arbitrary dependence on the generic function $f$ in both $\mathcal G$ and $T$ is exchanged by two scalar fields and scalar potential.
We then derive the JCs for matching between two different space-times across a separation hyper-surface $\Sigma$, assuming the matter sector to be described by an isotropic perfect fluid configuration. We take the general approach assuming the possibility of a thin-shell arising at $\Sigma$ between the two space-times. However, our results establish that, for the distribution formalism to be well-defined, thin-shells are not allowed to emerge in the general version of this theory. We thus obtain instead a complete set of JCs for a smooth matching at $\Sigma$ under the same conditions. The same results are then obtained in the scalar-tensor representation of the theory, thus emphasizing the equivalence between these two representations. Our results significantly constrain the possibility of developing models for alternative compact structures supported by thin-shells in $f(\mathcal{G},~T)$ gravity, e.g. gravastars and thin-shell wormholes, but provide a suitable framework for the search of models presenting a smooth matching at their surface, from which perfect fluid stars are possible examples.
\end{abstract}

\section{Introduction}
The preparation of appropriate JCs at surfaces of inconsistency is a core issue in both Newtonian and relativistic gravitational theories. There are two main types of these surfaces $(i)$ boundary surfaces, also known as shock waves, which are distinguished by a jump discontinuity in density, and $(ii)$ surface layers, which have an infinite density. The Schwarzschild and Oppenheimer-Snyder problems~\cite{oppenheimer1939continued} are well-known boundary surface problems in which the interior field of a static or collapsing star is linked to the exterior vacuum field, the Newtonian theory makes such problems simple to solve. The JCs are frequently required to establish suitable solutions between two space-times regions at $\Sigma$. An interesting concept in literature, one should address in any physical theory is how to treat discontinuous surfaces and their significant features. The JCs may connect the compact structure of the considered surface to the discontinuities of a physical quantity. In the study of Einstein's or modified field equations(MFEs) to find their solutions in the general theory of relativity(GR) or modified gravitational theories, one may come across the scenario in which a hyper-surface divides space-time into two different regions and two different co-ordinate systems serrate them. The gravitational JCs for singular hyper-surfaces establish the criteria between the two space-times which satisfy the hyper-surface and also correspond to a which satisfies the hyper-surface and corresponds to a comprehensive solution of gravitational field equations(GFEs). The scenario is relatively comparable in a gravitational framework, and the JCs may be determined for a hyper-surface that splits space-time into two zones $(i)$ interior and $(ii)$ exterior. Lanczos~\cite{lanczos1922bemerkung,lanczos1924flachenhafte} did pioneer work to propose gravitational JCs for singular hyper-surfaces in the background of GR, which is the best innovation of individual intellect. Later on, these JCs were used to find matching between different space times.

Lake~\cite{lake2017revisiting} studied the Darmois and Lichnerowicz JCs, and also revisit the equivalence of these JCs discussed in Gaussian normal coordinates. They showed that this equivalence of JCs did not extend to admissible type coordinates in which the metric across the hyper-surface and its $1st$-order derivatives represented the continuity at $\Sigma$. Sen~\cite{sen1924grenzbedingungen} picked up the interior and exterior as Minkowski, and Schwarzschild space-times, respectively, to continue his investigation for their JCs, and also discussed Buchdahl radius in the perspective of a thin-shell. Darmois~\cite{darmois1927memorial} devised a formulation of JCs that are based on extrinsic and intrinsic curvature, these conditions are a smooth match of extrinsic and intrinsic regions at $\Sigma$.
Synge and Brien~\cite{synge1952jump} discussed JCs and general hyper-surfaces formulated at discontinuities in GR. Furthermore, some significant JCs were originally discovered for non-null hyper-surfaces by Israel in~\cite{israel1966singular}, hence they were known as Israel's JCs which created a significant impact in the literature. The JCs for time-like case, null case, non-null case, space-like case, and generic hyper-surfaces were expressed in the form of the intrinsic as well as extrinsic curvatures of hyper-surfaces in~\cite{barrabes1989singular}.

Rosa and Picarra,~\cite{rosa2020existence} recently has demonstrated the stable behavior of relativistic spherically fluid with thin-shells and give some appropriate gravitational models. In the background of modified gravities, the relevance of JCs is obvious, because they're necessary for any explanation of matching between interior and exterior regions. Since different modified gravity theories will have their set of JCs, which will be determined from their equations of motion. For several modified gravity theories, the JCs are determined, some examples of such investigations are $f(R)$ gravitational theory, infinite derivative gravity, $f(R)$ gravity including torsion, Palatini $f(R)$, metric affine modified gravitational theory, Brane world models, and modified teleparallel theory of gravity etc.~\cite{senovilla2013junction,macias2002matching,de2014junction}.
Reina~\emph{et al.}~\cite{reina2016junction} constructed the JCs for the gravity theory with a Lagrangian that contains quadratic features in the curvature. The JCs have such a wide range of implementations because they can be utilized to extrapolate innovative solutions and thus introduce valuable insights into the underlying gravitational theory. Also investigation is done for irregularity factors for a spherically symmetric star in particular situations for both dissipative and non-dissipative domains.(see for further detail \cite{davis2003generalized,rosa2020stability,rosa2021double,
yousaf2021axially,bhatti2022dynamical}).

In modified gravities, like $f(R)$~\cite{buchdahl1970non} and $f(R,T)$ \cite{harko2011f} gravities which are widely utilized theories. Deruelle~\emph{et al.}~\cite{deruelle2008junction} used straightforward integration of the field equations to generalize the Israel JCs for modified gravity theories coupled to a scalar field. Olmo and Garcia~\cite{olmo2020junction} found the JCs for Palatini $f(R)$ gravitational theory by using a tensor distribution framework. Furthermore, they demonstrated the importance of these JCs by looking at the characteristics of astronomical surfaces in polytropic models and their significance in the physical realm. They utilized modified Chaplygin gas and found both stables as well as unstable regions, and also plotted their outcomes graphically. Roza \cite{rosa2021double} studied the double gravitational layer traversable wormholes solutions and also investigated the null energy condition in a quadratic-linear version of gravitational theory. Capozziello~\emph{et al.}~\cite{capozziello2012jeans} utilized the weak-field estimation to assess jean's dynamical stability including the collapse of collisionless self-gravitating compact structures, also obtained interacting collisionless Poisson and Boltzmann equations of motion from modified $f(R)$ cosmology.

Harko~\emph{et al.}~\cite{harko2011f} firstly generalized $f(R)$ gravitational theory to establish $f(R,T)$ gravity and also discussed requirements of this gravity. This is a notable version of the gravitational modified theory that incorporates all features of its fundamental gravities. Many researchers did their work under various relativistic geometries in the background of $f(R, T)$ cosmology. Sahoo~\emph{et al.}~\cite{sahoo2020wormhole} formulated a fresh composite structure of wormholes in the $f(R,~T)$ theory and also recommended that their purposed function fulfilled the criteria of wormhole geometry. Bhatti~\emph{et al.} also did work for some relativistic fluids in $f(R, T)$ theory to examine the stable behavior~\cite{bhatti2020stability,bhatti2021dynamical} by utilizing the adiabatic index approach for both minimally and non minimally coupled function. They also established the criteria for both stable and unstable compact fluid configurations in the Newtonian and post Newtonian realms under some essential physical constraints. Rosa and his collaborators~\cite{rosa2018wormholes} found wormhole solutions in extended hybrid-metric Palatini-gravitation. They also demonstrated that the solutions' main appeal is that the null energy conditions everywhere are enforced by the matter field, as well as at infinity, removing the necessity of exotic matter or dark matter.
The wormhole geometry, the existence of traversable as well as non-traversable wormholes by selecting two distinct shape functions, several cases by changing model parameters and checking the behavior
of wormhole geometry in the scenario of different gravities, it has been investigated in~\cite{godani2019traversable,mishra2021new,ghosh2021some,wheeler1962geometrodynamics,morris1988wormholes}.
Wang and Letelier~\cite{wang1995dynamical} investigated that there are two types of wormholes: static and dynamic. For the construction of static traversable wormholes and the exotic matter is required which
violates null energy conditions in the scenario of GR~\cite{hochberg1997geometric,hochberg1998dynamic}.

Rosa~\cite{rosa2021junction}, recently constructed the JCs for the matching of two space-times at a $\Sigma$, both in geometrical and dynamically equivalent scalar-tensor demonstrations. He established two different sets of JCs in formerly mentioned representations, and also showed the viability by considering some significant examples. Rosa and Lemos~\cite{rosa2021junctiona} discovered the junction criteria for an other extended gravity, and they also formulated two sets of JCs at $\Sigma$ for thin-shell and a smoothly match in both geometrical and dynamically equivalent scalar-tensor demonstrations by taking three different configurations to demonstrate the viability of all these JCs.
Barrabes and Hogan~\cite{barrabes2003singular} offered a comprehensive
framework for characterizing singular hyper-surfaces incorporating a Gauss-Bonnet term in Einstein's GR. They expressed JCs in a manner that is appropriate for any embedding and matter content, as well as coordinates selected separately on either side of the hyper-surface.
The modified Gauss-Bonnet theory is a fascinating gravity theory. A Gauss-Bonnet invariant is a mathematical function of the format $\mathcal{G}=R^2-4R^{i j}R_{i j}+R^{ijpq}R_{i j p q}$, in this mathematical function $R$ is Ricci scalar invariant which is the trace of $R_{i j}$, as well as $R_{i j}$ and $R_{i j p q}$ denote  Ricci, and the Riemann tensors, respectively. The Gauss-Bonnet mathematical formation is a $4$-dimensional topological invariant without gravitational field equations. When this combination is coupled with a scalar field or $f(\mathcal{G})$ a generic function is introduced to the Einstein-Hilbert action, it produces remarkable results in the same dimensions. Thus $f(\mathcal{G})$ gravity was presented by Nojiri and Odintsov which is commonly known as Gauss-Bonnet theory and also an alternative for dark energy~\cite{nojiri2005modified}.

Thus $f(\mathcal{G},~T)$ theory, like previous modified gravity theories, is a viable option for studying dark energy and shows the agreement with solar system requirements. In this framework, it is conceivable to explain the shift from decelerated to accelerated expansion, as well as the convergence of both early and late accelerating universe expansion.
Bhatti~\emph{et al.}~\cite{yousaf2020complexity,bhatti2021electromagnetic,bhatti2021role,yousaf2022f} studied the complexity of non-static compact spherical structure in the background of $f(\mathcal{G})$ gravity, also presented a new modified gravitational theory. They used Bel's procedure to orthogonally break down the Riemann tensor and formulated the complexity factor by using one of the scalar functions. Thermodynamics, cosmology, gravitational waves, the complexity and compact object astrophysics has all been explored using the different gravity
theory~\cite{khlopov2010composite,bamba2012dark,momeni2016generalized,moraes2016braneworld,
alves2016gravitational,yousaf2022hyperbolically,yousaf2022role}.
Capozziello~\emph{et al.} \cite{capozziello2003quintessence,capozziello2002curvature,capozziello2006dark} gave various novel perspectives on subjects such as quintessence and the fast expansion of our universe, contrasting them with observations and the role of dark matter. Bhatti \emph{et al.}~\cite{bhatti2020gravastars,yousaf2021electrically} discussed the structure of symmetric objects with and without electromagnetic effects in non-linear modified gravity theory and their comprehensive study helped to establish the alternative to black hole e.g. gravastars.

The paper will be organized as follows:
In the second section, we will present the general form of action integral and modified gravity theory in perfect fluid $f(\mathcal{G},~T)$ gravity. In the same section, we provide two representations of the $f(\mathcal{G},~T)$ theory that is dynamically equivalent by introducing two dynamical scalar fields as well as a scalar interaction potential. In the third section, we compute the JCs for matching without a thin-shell in the geometrical demonstration of the $f(\mathcal{G},~T)$ theory, and derive a set of JCs in the scenario of the same representation when $\mathcal{S}_{\mu\nu}$ disappears at $\Sigma$ i.e., is a smooth matching of $\mathbf{M}^{\pm}$ in the distribution formalism of this theory.
Since forth contains an established set of JCs for smoothly match at $\Sigma$ under same the perfect fluid configurations of considered gravitational theory, in a dynamically equivalent scalar-tensor illustration. In the last section, discussions and conclusions are briefly summed up.

\section{$f(\mathcal{G},~T)$ gravitational theory and field equations}

\subsection{Geometrical representation}
The action that describes the $f(\mathcal{G},~T)$ gravity can be obtained via a generalization of the Einstein-Hilbert action minimally coupled with an ordinary matter Lagrangian $\mathcal{L}_{\mathrm{m}}$ as
\begin{equation}\label{1i}
S=\frac{1}{2\kappa^2} \int_{\Omega}\sqrt{-g}\bigg[R+f\big(\mathcal{G},T\big)\bigg] \mathrm{d}^{4} x
+S_{\mathrm{m}},
\end{equation}
where $\Omega$ is a 4-dimensional space-time manifold on which a set of coordinates $x^i$ is defined, $S_{\mathrm{m}}=\int_{\Omega}\mathcal{L}_{\mathrm{m}}\sqrt{-g} d^{4}x$ is the matter action written in terms of a matter Lagrangian density $\mathcal{L}_{\mathrm{m}}$, $g$ is the determinant of the metric $g_{ij}$ written in terms of $x^i$ and with a signature $\left(+,-,-,-\right)$, $\kappa^2=8\pi G/c^4$ where $G$ is the gravitational constant and $c$ is the speed of light, and $f\left(\mathcal G,T\right)$ is an arbitrary function of the Gauss-Bonnet invariant $\mathcal G=R^2-4R_{ij}R^{ij}+R_{ijpq}R^{ijpq}$, in this mathematical function  $R=R^i_i$ is the Ricci scalar which is the trace of $R_{i j}$, as well as $R_{i j}$ and $R_{i j p q}$ denote  Ricci, and the Riemann tensors, respectively. And the trace $T$ of the stress-energy tensor $T_{ij}$, the latter is defined as a modification of the matter Lagrangian w.r.t. the $g_{ij}$ as
\begin{equation}\label{2i}
T_{i j}=-\frac{2}{\sqrt{-g}} \left(\frac{\delta\left(\sqrt{-g} \mathcal{L} _\mathrm{m}\right)}{\delta g^{i j}}\right)=g_{i j} \mathcal{L} _\mathrm{m}-\frac{2\partial \mathcal{L} _\mathrm{m}}{\partial g^{i j}}.
\end{equation}
Under the assumption that $T_{ij}$ depends solely on components of the metric $g_{ij}$ but not on its derivatives.
The MFEs of $f(\mathcal{G},~T)$ gravity may be derived by varying Eq.\eqref{1i} w.r.t. $g_{ij}$, from which one obtains
\begin{align}\nonumber&
R_{i j}-\frac{1}{2} g_{i j} R=\kappa^{2} T_{i j}+\frac{1}{2} g_{i j} f
+\mathcal G_{ij}f_{\mathcal{G}}-\left(T_{i j}+
\mathbf{\Theta}_{i j}\right) f_{T}
\\\label{4i}&-4\left(R_{i p j q} \nabla^{p}\nabla^{q}
- \frac{R}{2}\nabla_{i} \nabla_{j}-g_{i j}
R^{p q} \nabla_{p} \nabla_{q}+R_{i}^{p}
 \nabla_{j} \nabla_{p}-R_{i j} \nabla^{2}\right.
\left.
+\frac{R}{2}g_{i j} \nabla^{2}+R_{j}^{p} \nabla_{i} \nabla_{p}\right)f_{\mathcal{G}},
\end{align}
here we introduced the notation $f_\mathcal{G}\equiv\partial f/\partial \mathcal{G}$ and $f_T=\partial f/\partial T$ for the partial derivatives of the function $f\left(\mathcal G,T\right)$, $\nabla^2\equiv g^{ij}\nabla_i\nabla_j$ for the d'Alembert operator, where $\nabla_i$ are the 4-dimensional covariant derivatives expressed in the form of $g_{ij}$, the tensor $\mathcal G_{ij}$ is defined as
\begin{align}\label{5i}&
\mathcal G_{ij}\equiv- R R_{i j}+2 R_{i}^{p} R_{p j}+2 R_{i p j q} R^{p q}- R_{i}^{p q \delta} R_{j p q \delta},
\end{align}
and we have defined the auxiliary tensor $\mathbf{\Theta}_{ i j}$ in the form of variation of $T_{ij}$ w.r.t. $g_{ij}$ as
\begin{equation}\label{6i}
\mathbf{\Theta}_{ i j}=g^{pq}\frac{\delta T_{pq}}{\delta g^{ij}}.
\end{equation}
We will consider that the matter sector can be well characterized by a relativistic perfect fluid, i.e. $T_{ij}$ may be expressed as
\begin{align}\label{7i}&
T_{i j}=(\rho+P) u_{i} u_{j}-P g_{i j},
\end{align}
where $\rho$ and $P$ are the energy density and the isotropic surface pressure, respectively, and $u^i$ is the fluid 4-velocity vector. The normalization property is satisfied by the vector $u^i$ as $g_{ij}u^iu^j=1$, from which one obtains a matter Lagrangian density of the form $\mathcal{L}_{\mathrm{m}}=-P$. Consequently, Eq.\eqref{6i} yields
\begin{align}\label{8i}&
\mathbf{\Theta}_{i j}=-2 T_{i j}-P g_{i j}.
\end{align}
Introducing Eq.\eqref{8i} in Eq.\eqref{4i}, one obtains the MFEs for the isotropic perfect-fluid configuration as
\begin{align}\nonumber&
R_{i j}-\frac{R}{2} g_{i j} =\big(f_{T}+\kappa^{2}\big)T_{i j}+\mathcal G_{ij} f_{\mathcal{G}}+g_{i j}Pf_{T}+\frac{f}{2} g_{i j}
\\\label{9i}&-4\left(R_{i p j q} \nabla^{p}\nabla^{q}
- \frac{R}{2}\nabla_{i} \nabla_{j}-g_{i j}
R^{p q} \nabla_{p} \nabla_{q}+R_{i}^{p}
 \nabla_{j} \nabla_{p}-R_{i j} \nabla^{2}\right.
\left.+\frac{R}{2}g_{i j} \nabla^{2}+R_{j}^{p} \nabla_{i} \nabla_{p}\right)f_{\mathcal{G}}.
\end{align}
The differential terms on $f_\mathcal{G}$ may be expanded in the form of derivatives of $\mathcal G$ as well as $T$ with the use of the chain rule, as follows
\begin{align}\label{10i}
\partial_{i} f_{\mathcal{G}}=f_{\mathcal{G}\mathcal{G}} \partial_{i}\mathcal{G}+f_{\mathcal{G} T} \partial_{i} T,
\end{align}
\begin{align}\label{11i}&
\nabla_{i} \nabla_{j} f_{\mathcal{G}}=f_{\mathcal{G}\mathcal{G}} \nabla_{i} \nabla_{j}\mathcal{G}+f_{\mathcal{G} T} \nabla_{i} \nabla_{j}T+f_{\mathcal{G} T T} \partial_{i} T \partial_{j} T
+f_{\mathcal{G} \mathcal{G}\mathcal{G}} \partial_{i}\mathcal{G} \partial_{j}\mathcal{G}
+2 f_{\mathcal{G} \mathcal{G} T} \partial_{(i} \mathcal{G} \partial_{j)} T .
\end{align}
The fully extended field equations can now be conceived by plugging the expansions of Eqs.\eqref{10i} and \eqref{11i} into Eq.\eqref{9i}. However, we chose not to write the result explicitly because of its extensive nature.

\subsection{Dynamically equivalent scalar-tensor representation}
In the framework of modified gravities featuring extra scalar degrees of freedom in comparison to GR, it
is frequently effective to recast the geometrical representation of Eq.\eqref{1i} into a dynamically
equivalent scalar-tensor representation in which the additional scalar degrees of freedom are exchanged by scalar fields. In the scenario of $f\left(\mathcal{G}, T\right)$ gravity, the theory has two additional scalar degrees of freedom, representing the function $f$ which shows two arbitrary dependencies in the quantities $\mathcal G$ and $T$, and thus the scalar-tensor demonstration of this gravity can be conceived via the introduction of two scalar fields, say $\Phi$ and $\Psi$, as well as an interaction potential $\mathbb{V}\left(\Phi,\Psi\right)$. Let us start this transformation by introducing two auxiliary fields $\alpha$ and $\beta$ into Eq.\eqref{1i} which yield
\begin{align}\label{13i}&
S=\frac{1}{2 \kappa^{2}} \int_{\Omega} \sqrt{-g}\left\{R+f(\alpha, \beta)+\frac{\partial f}{\partial \alpha}\big(\mathcal{G}-\alpha\big)+
\frac{\partial f}{\partial \beta}\big(T-\beta\big)\right\} d^{4}x +S_{\mathrm{m}}.
\end{align}
The action in expression Eq.\eqref{13i} is explicitly reliant on three independent physical quantities: $g_{ij}$ as well as $\alpha$ and $\beta$. Consequently, the equations of motion for the auxiliary fields may be found by varying Eq.\eqref{13i} w.r.t. these fields, respectively, which yields
\begin{equation}\label{auxsys1}
\frac{\partial^2 f}{\partial \alpha^2}\bigg(\mathcal {G}-\alpha\bigg)+\frac{\partial^2 f}{\partial \alpha\partial \beta}\bigg(T-\beta\bigg)=0,
\end{equation}
\begin{equation}\label{auxsys2}
\frac{\partial^2 f}{\partial \alpha\partial \beta}\bigg(\mathcal {G}-\alpha\bigg)+\frac{\partial^2 f}{\partial \beta^2}\bigg(T-\beta\bigg)=0,
\end{equation}
where the function $f\left(\alpha,\beta\right)$ is assumed to satisfy the Schwartz theorem, i.e., its crossed partial derivatives are commutative. Equations \eqref{auxsys1} and \eqref{auxsys2} can be recast into a matrix form $\mathcal{A} \textbf{x}=0$ as
\begin{equation}\label{matrixeq}
\mathcal{A}\textbf{x}=\begin{pmatrix}
\frac{\partial^2 f}{\partial \alpha^2} & \frac{\partial^2 f}{\partial \alpha\partial \beta} \\
\frac{\partial^2 f}{\partial \alpha\partial \beta} & \frac{\partial^2 f}{\partial \beta^2}
\end{pmatrix}
\begin{pmatrix}
\mathcal{G}-\alpha \\
T-\beta
\end{pmatrix}=0.
\end{equation}
The solution of the matrix system in Eq.\eqref{matrixeq} will be unique if and only if $\mathcal{A}$ has non-zero determinant, i.e., if the following condition is satisfied:
\begin{equation}\label{condition}
\frac{\partial^2 f}{\partial \alpha^2}\frac{\partial^2 f}{\partial \beta^2}\neq \left(\frac{\partial^2 f}{\partial \alpha\partial \beta}\right)^2.
\end{equation}
If above mentioned Eq.\eqref{condition} is fulfilled, then the unique solution of established Eq.\eqref{matrixeq} is $\alpha=\mathcal{G}$ and $\beta=T$. When this solution is reintroduced into Eq.\eqref{13i}, then Eq.\eqref{1i} is recovered, confirming the equivalence of the two representations of the theory. Now scalar fields $\Phi$ and $\Psi$, as well as the interaction potential  $\mathbb{V}\left(\Phi,\Psi\right)$, may now be defined as follows
\begin{align}\label{14i}
&\Phi=f_\mathcal{G}\equiv\partial f/\partial \mathcal{G}, \quad \Psi=f_{T}\equiv\partial f/\partial T, \\\label{15i}
&\mathbb{V}(\Phi,~\Psi)=\alpha \Phi+\beta \Psi-f(\alpha, \beta),
\end{align}
which upon replacement into the action in Eq.\eqref{13i} yields the modified Einstein Hilbert action of considered representation of the $f\left(\mathcal{G},T\right)$ gravity in the form
\begin{align}\label{17i}
S&=\frac{1}{2} \int_{\Omega} \frac{\sqrt{-g}}{\kappa^{2}}\left\{R+\Phi\mathcal{G}+\Psi T-\mathbb{V}\big(\Phi, \Psi\big)\right\}d^{4} x
+S_{\mathrm{m}}.
\end{align}
Since then, Eq.\eqref{17i} is now explicitly reliant on three independent quantities: the metric $g_{ij}$, as well as $\Phi$ and $\Psi$. By varying Eq.\eqref{17i} w.r.t. the metric $g_{ij}$, one may get the MFEs in the scalar-tensor demonstration, whereas the equations of motion for the scalar fields are conceived via the variation w.r.t. $\Phi$ and $\Psi$. These equations take the forms
\begin{align}\nonumber&
R_{i j}-\frac{1}{2}(R+\Phi\mathcal{G}+\Psi T-\mathbb{V})g_{i j}+\Phi\mathcal{G}_{i j}
+4\left(R_{i p j q} \nabla^{p}\nabla^{q}
- \frac{R}{2}\nabla_{i} \nabla_{j}-g_{i j}
R^{p q} \nabla_{p} \nabla_{q}+R_{i}^{p}
 \nabla_{j} \nabla_{p}\right.
\\ \label{18i} &\left.-R_{i j} \nabla^{2}
+\frac{R}{2}g_{i j} \nabla^{2}+R_{j}^{p} \nabla_{i} \nabla_{p}\right)\Phi
=\kappa^{2} T_{i j}
-\left(T_{i j}+\Theta_{i j}\right)\Psi,
\end{align}
\begin{align}\label{19iG}&
\mathbb{V}_{\Phi}=\mathcal{G},\\\label{19iT}&
\mathbb{V}_{\Psi}=T,
\end{align}
respectively, where we have introduced the notation $\mathbb{V}_{\Phi}\equiv \frac{\partial \mathbb{V}}{\partial\Phi}$ and $\mathbb{V}_{\Psi}\equiv \frac{\partial \mathbb{V}}{\partial\Psi}$ for the partial derivatives of interaction potential. It should be noted that Eq.\eqref{18i} may be produced directly from Eq.\eqref{4i} by introducing the definitions in expressions Eqs.\eqref{14i} and \eqref{15i} and the use of the following geometrical identity valid only in four dimensions
$$
R R_{ij}-2R_i^pR_{pj}-2R_{ipjq}R^{pq}+R_i^{\ pqr}R_{jpqr}=\mathcal{G}_{ij}.
$$
Following the same procedure as for the geometrical representation, i.e., assuming that the matter sector is characterized by an isotropic perfect fluid, which implies that the stress-energy tensor is Eq.\eqref{7i}, which implies that the matter Lagrangian is $\mathcal{L}_{\mathrm{m}}=-P$ as well as the auxiliary tensor $\mathbf{\Theta}_{i j}$ is given in the form of Eq.\eqref{8i}, Eq.\eqref{18i} can be recast in a more convenient manner
\begin{align}\nonumber&
R_{i j}-\frac{1}{2}(R+\Phi\mathcal{G}+\Psi T-\mathbb{V})g_{i j}
+\Phi\mathcal{G}_{i j}+4\left(R_{i p j q} \nabla^{p}\nabla^{q}
- \frac{R}{2}\nabla_{i} \nabla_{j}-g_{i j}
R^{p q} \nabla_{p} \nabla_{q}+R_{i}^{p}
 \nabla_{j} \nabla_{p}\right.
\\\label{20i}&\left.-R_{i j} \nabla^{2}
+\frac{R}{2}g_{i j} \nabla^{2}+R_{j}^{p} \nabla_{i} \nabla_{p}\right)\Phi
=\left(\kappa^{2}+\Psi\right)T_{i j}+Pg_{i j}\Psi.
\end{align}

\section{Junction conditions of the $f(\mathcal{G},~T)$ gravity}

\subsection{Notation and assumptions}
\begin{figure}[h!]\centering
\epsfig{file=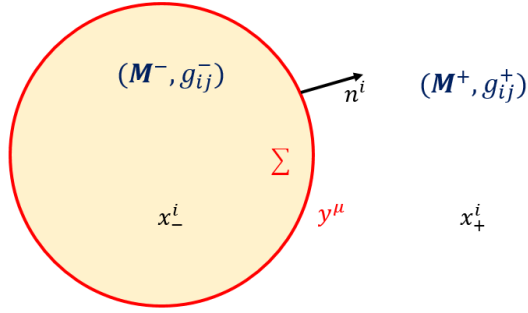,width=0.5\linewidth}
\caption{In this schematic representation a spherical hyper-surface $\Sigma$ in red divides the space-time $\mathbf{M}$ into two sections $\mathbf{M}^{\pm}$. The the coordinate systems $x^i_\pm$ defined in $\mathbf{M}^{\pm}$, respectively, where as the coordinate systems specified at the hyper-surface $\Sigma$ are $y^{\mu}$, and the spacelike unit vector normal to $\Sigma$ is $n^i$.}\label{1}
\end{figure}

Take the space-time to be given by a manifold $(\mathbf{M}, g_{i j})$ which can be divided into two distinct regions, a region $\mathbf{M}^{+}$ described by a metric $g_{ij}^+$ written in terms of a coordinate system $x^i_+$, and a region $\mathbf{M}^{-}$ described by a metric $g_{ij}^-$ written in terms of a coordinate system $x^i_-$. These two regions are separated by a hyper-surface $\Sigma$, on both sides of which we specify a coordinate system $y^\mu$, the direction perpendicular to $\Sigma$ is excluded from Greek indices. The projection vectors from the four-dimensional manifold $\mathbf{M}$ to the three-dimensional hyper-surface $\Sigma$ may then be expressed as follows: $e^i_\mu=\partial x^i/\partial y^\mu$, and we define the normal unit vector orthogonal to the hyper-surface and pointing from the region $\mathbf{M}^-$ to the region $\mathbf{M}^+$ as $n^i$, which implies that the result $e^i_\mu n_i=0$ holds. Schematic construction of our considered scenario can be found in Figure \ref{1}. The displacement from $\Sigma$ along the geodesic congruence generated by the normal vector $n^i$ is parameterized by an affine parameter $l$, i.e., one can write
\begin{equation}\label{22i}
dx^i=n^i dl, \qquad n_i=\varepsilon\partial_i l,
\end{equation}
where $\varepsilon\equiv n^in_i$ can be either $1$ or $-1$ depending on the normal vector being spacelike or timelike, respectively. Furthermore, we chose the affine parameter $l$ in such a way to guarantee that $l>0$ in the region $\mathbf{M}^+$, $l<0$ in the region $\mathbf{M}^-$, and $l=0$ at the hyper-surface $\Sigma$.

A suitable mathematical approach to deal with these situations where the space-time can be divided into different regions is the distribution formalism. In this formalism, any regular quantity $Y$ defined in the entire space-time manifold $\mathbf{M}$ may be expressed in terms of the Heaviside distribution function $\mathbf{\Theta}(l)$ i.e.,
\begin{align}\label{21i}&
Y=Y^{+} \mathbf{\Theta}(l)+Y^{-}\mathbf{\Theta}(-l),
\end{align}
where $Y^\pm$ denote the forms of the quantity $Y$ in the regions $\mathbf{M}^\pm$, respectively. The Heaviside function is thus defined as $0$ in the region $l<0$, as $1$ in the region $l>0$, and as $\frac{1}{2}$ at the point $l=0$, which corresponds to $\Sigma$. We noted in this formulism the derivative of the Heaviside function and the following properties:
\begin{equation}\label{Basicproperties}
\mathbf{\Theta}^2(l)=\mathbf{\Theta}(l), ~~\frac{d}{dl}\mathbf{\Theta}(l)=\delta\left(l\right),~~\mathbf{\Theta}(l)\mathbf{\Theta}(-l)=0,
\end{equation}
where $\delta\left(l\right)$ is the Dirac-delta distribution function~\cite{eric2007relativists}. We also noted that the product $\mathbf{\Theta}(l)\delta\left(l\right)$ is not defined as a distribution. Furthermore, the jump of the quantity $Y$ across the hyper-surface $\Sigma$ as well as the value of the quantity $Y$ at the hyper-surface $\Sigma$ are defined
as
\begin{equation}\label{23i}
[Y]=Y^{+}|_{\Sigma}-Y^{-}|_{\Sigma}.
\end{equation}
\begin{equation}
Y^\Sigma=\frac{1}{2}\left(Y^{+}|_{\Sigma}+Y^{-}|_{\Sigma}\right).
\end{equation}
If a quantity $Y$ is continuous, then its jump satisfies $\left[Y\right]=0$ and one obtains $Y^{+}|_{\Sigma}=Y^{-}|_{\Sigma}=Y^\Sigma$. Thus, by construction, the normal vector $n^i$ and the projection vectors $e^i_\mu$ satisfy the property $\left[n^{i}\right]=\left[e_{\mu}^{i}\right]=0$.

\subsection{Junction conditions in the geometrical representation}

We now have introduced all the necessary tools for the analysis that follows. Let us now focus on the general scenario for which a thin-shell is allowed to emerge at the separation hyper-surface $\Sigma$ and use the distribution formalism described above to deduce the corresponding JCs of the theory. For the line element to be fully specified on the entire space-time manifold $\mathbf{M}$ and, in particular, on both sides of the hyper-surface $\Sigma$, in the distribution formalism, we express $g_{ij}$ as
\begin{align}\label{25i}
g_{i j}=g_{i j}^{+} \mathbf{\Theta}(l)+g_{i j}^{-} \mathbf{\Theta}(-l).
\end{align}
The partial derivatives of $g_{i j}$ are as follows:
$$\partial_{w} g_{i j}=\partial_{w} g_{i j}^{+} \mathbf{\Theta}(l)+\partial_{w} g_{i j}^{-} \mathbf{\Theta}(-l)+ \varepsilon\left[g_{i j}\right] n_{w} \delta(l).$$
In the derivatives of the metric, there appears to be a component proportional to $\delta(l)$, which is problematic. Indeed, when one tries to establish the Christoffel symbols $\Gamma^p_{ij}$ related to the metric $g_{ij}$, these terms result in the occurrence of products of the type $\mathbf{\Theta}(l)\delta(l)$ in $\Gamma^p _{ij}$, which will result in terms proportional to $\delta\left(l\right)$ at $\Sigma$ since $\mathbf{\Theta}(0)=\frac{1}{2}$. These terms are problematic when one defines the Riemann tensor $R^i_{\ jpq}$, as it depends on products of Christoffel symbols and thus it depends on terms of the type $\delta^2(l)$, that is singular in considered distribution formalism. To avoid the existence of these highly problematic products, the continuity of the metric $g_{ij}$ must be imposed i.e.,
\begin{align}\label{27i}&
\left[g_{i j}\right]=0.
\end{align}
This expression applicable to the coordinate system $x^{i}$ only. Furthermore, we can simply convert this into a coordinate invariant statement: $0=[g_{i j}] e_{\mu}^{i} e_{\nu}^{j}=[g_{i j} e_{\mu}^{i} e_{\nu}^{j}];$ this last step comes as a result of the property $\left[n^{i}\right]=\left[e_{\mu}^{i}\right]=0$.
The metric intrinsic to the hyper-surface $\Sigma$ is produced by confining the line element to displacements restricted to the hyper-surface. Consequently the parametric equations $x^{i}=x^{i}\left(y^{\mu}\right)$, we have the vectors
$e^i_\mu=\frac{\partial x^i}{\partial y^\mu}$
are tangent to curves contained in $\Sigma$. Now, for displacements within separation hyper-surface $\Sigma$ we have
$$
\begin{aligned}
\mathrm{d} s^{2} &=g_{i j} \mathrm{d} x^{i} \mathrm{~d} x^{j} \\
&=g_{i j}\left(\frac{\partial x^i}{\partial y^\mu} \mathrm{~d} y^{\mu}\right)\left(\frac{\partial x^i}{\partial y^\nu} \mathrm{~d} y^{\nu}\right) \\
&=\mathbf{h}_{\mu \nu} \mathrm{~d} y^{\mu} \mathrm{~d} y^{\nu}
\end{aligned}
$$
where $\mathbf{h}_{\mu \nu}=g_{i j} e_{\mu}^{i} e_{\nu}^{j}$
is $1st$ fundamental form, of the hyper-surface $\Sigma$. We will refer to such objects as three-tensors.
The argument above can be extended to the induced metric $\mathbf{h}_{\mu \nu}=g_{ij}e^i_\mu e^j_\nu$ at the hyper-surface $\Sigma$. The induced metric from the exterior can be written as $\mathbf{h}_{\mu \nu}^{+}=g_{i j}^{+} e_{\mu}^{i} e_{\nu}^{j}$, whereas the induced metric from the interior is $\mathbf{h}_{\mu \nu}^{-}=g_{i j}^{-} e_{\mu}^{i} e_{\nu}^{j}$. Since we have established that $\left[g_{i j}\right]=0$, this implies that $\mathbf{h}_{\mu \nu}^{+}-\mathbf{h}_{\mu \nu}^{-}=0$ and we obtain the $1st$ JC of the theory in the form
\begin{align}\label{29i}
\left[\mathbf{h}_{\mu \nu}\right]=0 .
\end{align}
The JC obtained in Eq.\eqref{29i} corresponds also to the $1st$ JC of GR, and it is true for a wide range of metric theories of gravity, including the $f(\mathcal{G},T)$ we study in this work. Following Eq.\eqref{27i}, the partial derivatives of $g_{ij}$ take the regular form
\begin{align}\label{30i}
\partial_{w} g_{i j}=\partial_{w} g_{i j}^{+} \mathbf{\Theta}(l)+\partial_{w} g_{i j}^{-} \mathbf{\Theta}(-l),
\end{align}
and the Christoffel symbols are well-defined in the distribution formalism. One can now proceed to construct the remaining geometrical quantities in the distribution formalism, namely the Riemann tensor $R_{\ p i j}^{q}=\partial_i \Gamma_{p j}^{q}-\partial_j  \Gamma_{p i}^{q}+\Gamma_{r i}^{q} \Gamma_{p i}^{r}-\Gamma_{r j}^{q} \Gamma_{p i}^{r}$, the Ricci tensor $R_{ij}=R^p_{\ ipj}$, and the Ricci scalar $R=R^i_i$. These quantities take the forms
\begin{equation}\label{rabcddist}
R_{ijpq}=R^{+}_{ijpq} \mathbf{\Theta}(l)+R^{-}_{ijpq} \mathbf{\Theta}(-l)+\bar R_{ijpq}\delta(l),
\end{equation}
\begin{equation}\label{rabdist}
R_{ij}=R^{+}_{ij} \mathbf{\Theta}(l)+R^{-}_{ij} \mathbf{\Theta}(-l)+\bar R_{ij}\delta(l),
\end{equation}
\begin{equation}\label{rdist}
R=R^{+} \mathbf{\Theta}(l)+R^{-} \mathbf{\Theta}(-l)+\bar R\delta(l),
\end{equation}
where the quantities $\bar R_{i j p q}$, $\bar R_{i j}$ and $\bar R$ are given by
\begin{equation}\label{rabcdsing}
\bar R_{i j p q}=\left[\mathcal K_{\mu\nu}\right]\left(e^{\mu}_{i}e^{\nu}_{q} n_{j} n_{p}-e^\mu_{i}e^\nu_{p} n_{j} n_{q}-e^\mu_{j}e^{\nu}_{q} n_{i} n_{p}+e^{\mu}_{j}e^{\nu}_{p} n_{i} n_{q}\right),
\end{equation}
\begin{equation}\label{rabsing}
\bar R_{ij}=-\left(\varepsilon \left[\mathcal K_{\mu\nu}\right]e^\mu_i e^\nu_j+n_i n_j \left[\mathcal K\right]\right),
\end{equation}
\begin{equation}\label{rsing}
\bar R=-2\varepsilon\left[\mathcal K\right],
\end{equation}
here $\mathcal K_{\mu\nu}=e^i_\mu e^j_\nu \nabla_i n_j$ indicates the extrinsic curvature of $\Sigma$, also known as the second fundamental form, and $\mathcal K=\mathcal{K}_\mu^\mu$ is its trace. From the expressions above it is clear that the singular parts of the Riemann and the Ricci tensors will disappear if and only if the $\left[\mathcal K_{\mu\nu}\right]=0$, i.e., the jump of the extrinsic curvature disappears, and the singular portion of the Ricci scalar will vanish if and only if the jump of the trace of the extrinsic curvature disappears, i.e., $\left[\mathcal K\right]=0$.

To proceed with our analysis, we also need to find the explicit form of the Gauss-Bonnet invariant $\mathcal{G}=R^2-4R^{i j}R_{i j}+R^{ijpq}R_{i j p q}$ in the distribution formalism. Using the results obtained in Eqs.\eqref{rabcddist} to \eqref{rdist} into the expression for $\mathcal{G}$,
one expects the Gauss-Bonnet invariant to take a form of the type:
\begin{equation}\label{37i1}
\mathcal G=\mathcal{G}^{+} \mathbf{\Theta}(l)+\mathcal{G}^{-} \mathbf{\Theta}(-l)+\bar{\mathcal{G}}\delta\left(l\right)
+\hat{\mathcal{G}}\delta^2\left(l\right),
\end{equation}
because the terms corresponding to $\delta\left(l\right)$ are present in $R_{ijpq}$, $R_{ij}$ and $R$.  However, this calculation shows that the term proportional to $\delta\left(l\right)$, i.e.,
$$\hat{\mathcal G}=\bar R^2-4\bar R_{ij}\bar R^{ij}+\bar R_{ijpq}\bar R^{ijpq},$$
vanishes identically. Using this fact one can thus write the Gauss-Bonnet invariant in the distribution formalism as
\begin{equation}\label{37i}
\mathcal{G}=\mathcal{G}^{+} \mathbf{\Theta}(l)+\mathcal{G}^{-} \mathbf{\Theta}(-l)+\bar{\mathcal G}\delta\left(l\right),
\end{equation}
here $\bar{\mathcal{G}}$ can be expressed in the form of the geometrical quantities given previously as
\begin{equation}
\bar{\mathcal G}=R^\Sigma \bar R -4 R_{ij}^\Sigma \bar R^{ij}+R^\Sigma_{ijpq} \bar R^{ijpq}.
\end{equation}
Although the term $\bar{\mathcal G}$ is not problematic by itself, one must note that we are working in the $f\left(\mathcal G,~T\right)$ gravitational theory. This shows arbitrary dependence of the function $f$ in $\mathcal G$, one can expect that this function admits a Taylor series expansion in $\mathcal G$, which will feature power-laws of $\mathcal G$, thus giving rise to singular terms of the form $\delta^2\left(l\right)$. To prevent the occurrence of these terms, one must impose $\bar{\mathcal G}=0$. According to Eqs.\eqref{rabcdsing} to \eqref{rsing}, one verifies that for $\bar{\mathcal G}$ to vanish it is necessary to impose that $\left[\mathcal K_{\mu\nu}\right]=0$. One thus obtains the second JC in the form
\begin{equation}
\left[\mathcal K_{\mu\nu}\right]=0.
\end{equation}
This is a highly restrictive JC that is not present in GR or any of the other previously extensions of GR except for the particular case of smoothly match, i.e., matching without thin-shells, and it is linked to an arbitrary dependence of the action in $\mathcal G$. This condition implies that all the quantities $\bar R_{ijpq}$, $\bar R_{ij}$ and $\bar R$ vanish, consequently all the terms proportional to $\delta\left(l\right)$ in Eqs.\eqref{rabcddist} to \eqref{rdist} vanish, and these geometrical quantities become regular at $\Sigma$. We thus obtain the regular forms
\begin{align}\label{rabcdreg}
R_{ijpq}&=R^{+}_{ijpq} \mathbf{\Theta}(l)+R^{-}_{ijpq} \mathbf{\Theta}(-l)
,
\\\label{rabreg}
R_{ij}&=R^{+}_{ij} \mathbf{\Theta}(l)+R^{-}_{ij} \mathbf{\Theta}(-l)
,
\\\label{rreg}
R&=R^{+} \mathbf{\Theta}(l)+R^{-} \mathbf{\Theta}(-l)
,
\\\label{gbreg}
\mathcal{G}&=\mathcal{G}^{+} \mathbf{\Theta}(l)+\mathcal{G}^{-} \mathbf{\Theta}(-l).
\end{align}

Now consider the differential terms in the MFEs, see Eq.\eqref{9i}. Following the chain rule in Eq.\eqref{11i}, one verifies that these $2nd$-order differential terms in $f_\mathcal{G}$ can be expressed in the form of $1st$ and $2nd$-order differential terms of $\mathcal{G}$, and thus we must compute these derivatives in the considered formalism. Applying the $1st$-order derivative of Eq.\eqref{gbreg} one gets the form:
\begin{align}\label{41i1}
\partial_{i}\mathcal{G}=\partial_{i} \mathcal{G}^{+}\mathbf{\Theta}(l)+\partial_{i} \mathcal{G}^{-} \mathbf{\Theta}(-l)+\varepsilon[\mathcal{G}] n_{i} \delta(l).
\end{align}
According to Eq.\eqref{11i}, the $2nd$-order differential terms in $f_\mathcal{G}$ contain products of the form $\partial_i\mathcal{G}\partial_j\mathcal{G}$ which, according to the result just derived for $\partial_i\mathcal{G}$, will feature singular products of the form $\delta^2(l)$. To prevent the occurrence of these terms, one verifies that the Gauss-Bonnet invariant must be continuous across $\Sigma$, i.e., we can write the $3rd$ JC of the theory as
\begin{align}\label{41i}&
[\mathcal{G}]=0 .
\end{align}
This JC is also absent in GR and is linked to an arbitrary dependency of the action in the Gauss-Bonnet invariant $\mathcal{G}$, and thus it is expected to arise also in other similar theories like e.g. the simpler $f\left(\mathcal{G}\right)$ theory. Under this condition, the partial derivatives of $\mathcal{G}$ simplify to
\begin{align}\label{42i}&
\partial_{i} \mathcal{G}=\partial_{i} \mathcal{G}^{+}\mathbf{\Theta}(l)+\partial_{i} \mathcal{G}^{-} \mathbf{\Theta}(-l).
\end{align}
One can now take the $2nd$-order covariant derivatives of the Gauss-Bonnet invariant $\mathcal{G}$, i.e., which now take the form
\begin{align}\label{43i}&
\nabla_{i} \nabla_{j} \mathcal{G}=\nabla_{i} \nabla_{j} \mathcal{G}_{+}\mathbf{\Theta}(+l)+\nabla_{i}\nabla_{j} \mathcal{G}_{-}\mathbf{\Theta}(-l)+\varepsilon n_{i}\left[\partial_{j} \mathcal{G}\right] \delta(l).
\end{align}
Let us investigate the matter section of the MFEs. In the previous calculations, we have shown the $2nd$-order covariant derivatives of $\mathcal{G}$ in expression Eq.\eqref{43i}, feature terms proportional to $\delta\left(l\right)$ which linked to the occurrence of a perfect fluid thin-shell at $\Sigma$. In this representation $\mathcal{S}_{ij}$ denoting the stress-energy tensor of the thin-shell, one can write the stress-energy tensor $T_{ij}$ in the distribution formalism as
\begin{align}\label{46i}&
T_{i j}=T_{i j}^{+} \mathbf{\Theta}(l)+T_{i j}^{-}\mathbf{\Theta}(-l)+\mathcal{S}_{i j}\delta(l).
\end{align}
The quantity $\mathcal{S}_{ij}$ is a four-dimensional tensor representing the thin-shell. This quantity can be expressed as a three-dimensional $\mathcal{S}_{\mu\nu}$ tensor at $\Sigma$ via the projection
\begin{align}\label{47i}&
\mathcal{S}_{i j}=\mathcal{S}_{\mu \nu} e_{i}^{\mu} e_{j}^{\nu}.
\end{align}
The trace $T$ can thus be expressed in the distribution formalism as:
$$
T=T^{+} \mathbf{\Theta}(l)+T^{-} \mathbf{\Theta}(-l)+ \mathcal{S}\delta(l),
$$
here $\mathcal{S}=\mathcal{S}^i_i$ is the trace of the stress-energy tensor of the thin-shell. Now, since the function $f\left(\mathcal{G}, T\right)$ features an arbitrary dependence on the quantity $T$, one can expect in general that this function admits a Taylor series expansion in $T$, which will feature power-laws of this quantity. These power-laws will then feature the usual singular terms of the form $\delta^2(l)$ that must be removed to preserve the definiteness of the function. Thus, one concludes that the trace $\mathcal{S}$ must vanish, i.e., the $4th$ JC takes the form
\begin{align}\label{49i}&
\mathcal{S}=0.
\end{align}
This JC is related to the arbitrary reliance of the action in $T$ and it is also featured in other well-know theories e.g. $f\left(T\right)$ gravity and both the metric and the Palatini formulation of $f\left(R,T\right)$ gravity. Following this condition, the trace $T$ simplifies to
\begin{align}\label{50i}&
T=T^{+} \mathbf{\Theta}(l)+T^{-} \mathbf{\Theta}(-l).
\end{align}
Turning now to the differential terms in the MFEs, and similarly to what happens to the differential terms in $\mathcal{G}$, the $2nd$-order covariant derivatives of $f_\mathcal{G}$ can be rewritten in terms of $1st$ and $2nd$ order derivatives of $T$ via the chain rule in Eq.\eqref{11i}. Applying the $1st$-derivatives of $T$ one obtains $\partial_{i} T=\partial_{i} T^{+} \mathbf{\Theta}(l)+\partial_{i} T^{-} \mathbf{\Theta}(-l)+\varepsilon[T] n_{i} \delta(l)$. In Eq.\eqref{11i} one verifies that products of the form $\partial_i T\partial_j T$ are present in the $2nd$-order differential terms of $f_{\mathcal{G}}$, which give rise to singular products of the form $\delta^2(l)$. This products can be avoided by imposing the continuity of the trace $T$, i.e., the $5th$ JC takes the form
\begin{align}\label{52i}&
[T]=0.
\end{align}
This JC is also associated with the arbitrary dependence of the action in $T$ and it is also present e.g. in both formulations of $f\left(R,T\right)$ gravity (i.e., metric and Palatini). Under this condition, the $1st$ derivatives of $T$ simplify to
\begin{align}\label{53i}&
\partial_{i} T=\partial_{i} T^{+} \mathbf{\Theta}(l)+\partial_{i} T^{-} \mathbf{\Theta}(-l) .
\end{align}
Consequently, applying the $2nd$-order covariant derivatives of $T$ one obtains the form
\begin{align}\label{54i}&
\nabla_{i} \nabla_{j} T =\nabla_{i} \nabla_{j} T_{+}\mathbf{\Theta}(l)+\nabla_{i} \nabla_{j} T_{-} \mathbf{\Theta}(-l)+\varepsilon n_{i}\left[\partial_{j} T\right] \delta(l).
\end{align}
Now we have constructed all the necessary quantities in distribution formalism to obtain the singular part of the field equations and derive an explicit expression for the stress-energy tensor of the thin-shell in terms of the geometrical quantities. To do so, we introduce Eqs.\eqref{25i}, \eqref{rabcdreg}, \eqref{rabreg}, \eqref{rreg}, \eqref{gbreg}, \eqref{42i}, \eqref{43i}, \eqref{46i}, \eqref{50i}, \eqref{53i}, and \eqref{54i}, into the MFEs in Eq.\eqref{9i}, we project the result onto $\Sigma$ by utilizing the projection vectors $e^i_\mu$, and discard all non-singular terms. The result is as follows
\begin{eqnarray}\nonumber
\left(k^2+f_{T}\right)\mathcal{S}_{\mu \nu}&=& 4\varepsilon\left(e^i_\mu e^j_\nu R_{\ i\ j}^{p\ q}-R^{pq}\textbf{h}_{\mu\nu}\right)n_p\left(f_{\mathcal{G}\mathcal{G}}\left[\partial_{q} \mathcal{G}\right]+f_{\mathcal{G} T}\left[\partial_{q} T\right]\right)
\\\label{58i}&+&4\varepsilon \big(e_{\mu}^{i}e_{\nu}^{j}R_{ij}-\frac{1}{2}R\mathbf{h}_{\mu \nu} \big)n^{p}\left(f_{\mathcal{G}\mathcal{G}}\left[\partial_{p} \mathcal{G}\right]+f_{\mathcal{G} T}\left[\partial_{p} T\right]\right).
\end{eqnarray}
We can now use the trace of Eq.\eqref{58i} and the fact that $\mathcal{S}=0$ to recast the $4th$ JC in a more convenient form as
$$f_{\mathcal{G}\mathcal{G}}\left[\partial_{p} \mathcal{G}\right]+f_{\mathcal{G}T}\left[\partial_{p} T\right]=0.$$
Consequently, the right-hand side of Eq.\eqref{58i} disappears identically, which implies that $\mathcal S_{\mu\nu}=0$. This result indicates that under this formalism, thin-shells can not exist in the $f\left(\mathcal G,T\right)$ gravitational theory. This is a very restrictive result that forces all the junctions between two different space-times at a given hyper-surface $\Sigma$ to be smooth in this theory. The whole set of JCs for the geometrical representation of the $f\left(\mathcal{G},T\right)$ theory may thus be written as
\begin{equation}
\centering
\left.\begin{split}
\left[\mathbf{h}_{\mu \nu}\right]=0,\\
[\mathcal{K}_{\mu\nu}]=0,\\
[\mathcal{G}]=0,\\
[T]=0,\\
f_{\mathcal{G}\mathcal{G}}\left[\partial_{w} \mathcal{G}\right]+f_{\mathcal{G}T}\left[\partial_{w} T\right]=0.\\
\end{split}~~\right\}
 \end{equation}

The set of JCs is thus composed of five equations.

\subsection{Junction conditions in the scalar-tensor representation}

Now we will study the scalar-tensor representation of the $f(\mathcal{G},~T)$ gravitational theory, which was derived in the previous section. The method followed in this representation to obtain the $1st$ JC is identical to that used in the geometric representation, i.e., in the distribution formalism, we $1st$ write the metric $g_{ij}$ as
\begin{align}\label{1a}
g_{i j}=g_{i j}^{+} \mathbf{\Theta}(l)+g_{i j}^{-} \mathbf{\Theta}(-l).
\end{align}
Again, the appearance of a term proportional to $\delta(l)$ in the derivatives of the metric is problematic. The Christoffel symbols $\Gamma^p_{ij}$ linked to $g_{ij}$ will feature of products of the type $\mathbf{\Theta}(l)\delta(l)$, which will result in terms proportional to $\delta\left(l\right)$ at $\Sigma$ since $\mathbf{\Theta}(0)=\frac{1}{2}$. When one defines the Riemann tensor $R^i_{\ jpq}$, this tensor will depend on products of the form $\delta^2(l)$, as it depends on products of Christoffel symbols. These products are singular in the distribution formalism, and must also be avoided by enforcing the continuity of $g {ij}$, i.e., $\left[g_{i j}\right]=0$. Additionally, because $g_{i j}$ induces a metric on $\Sigma$ defined as $\mathbf{h}_{\mu \nu}=e_{\mu}^{i} e_{\nu}^{j} g_{i j}$, for the induced metric $\mathbf{h}_{\mu \nu}$, the same result must hold, i.e., the $1st$ JC takes the form
\begin{align}\label{2a}
\left[\mathbf{h}_{\mu \nu}\right]=0.
\end{align}
This is the same result as previously obtained in Eq.\eqref{29i} and does not depend on the representation of the theory used as a framework. The partial derivatives of $g_{i j}$ thus get the form $\partial_{w} g_{i j}=\partial_{w} g_{i j}^{+} \mathbf{\Theta}(l)+\partial_{w} g_{i j}^{-} \mathbf{\Theta}(-l)$, and consequently the Christoffel symbols for the metric $g_{i j}$ become well-defined in the distribution formalism. This allows for other three physical quantities: the Riemann tensor $R_{i j p q}$, Ricci tensor $R_{i j}$ and the Ricci scalar $R$ to be computed as well in the distribution formalism, which is given by the same forms as in the geometrical representation, i.e.,
\begin{equation}\label{3a}
R_{ijpq}=R^{+}_{ijpq} \mathbf{\Theta}(l)+R^{-}_{ijpq} \mathbf{\Theta}(-l)+\bar R_{ijpq}\delta(l),
\end{equation}
\begin{equation}\label{4a}
R_{ij}=R^{+}_{ij} \mathbf{\Theta}(l)+R^{-}_{ij} \mathbf{\Theta}(-l)+\bar R_{ij}\delta(l),
\end{equation}
\begin{equation}\label{5a}
R=R^{+} \mathbf{\Theta}(l)+R^{-} \mathbf{\Theta}(-l)+\bar R\delta(l).
\end{equation}
where the quantities $\bar R_{ijpq}$, $\bar R_{ij}$ and $\bar R$ are given by
\begin{equation}\label{3aa}
\bar R_{ijpq}=\left[\mathcal K_{\mu\nu}\right]\left(e^{\mu}_{i}e^{\nu}_{q} n_{j} n_{p}-e^\mu_{i}e^\nu_{p} n_{j} n_{q}-e^\mu_{j}e^{\nu}_{q} n_{i} n_{p}+e^{\mu}_{j}e^{\nu}_{p} n_{i} n_{q}\right),
\end{equation}
\begin{equation}\label{4aa}
\bar R_{ij}=-\left(\varepsilon \left[\mathcal K_{\mu\nu}\right]e^\mu_i e^\nu_j+n_i n_j \left[\mathcal K\right]\right),
\end{equation}
\begin{equation}\label{5aa}
\bar R=-2\varepsilon\left[\mathcal K\right],
\end{equation}
with $\mathcal{K}_{\mu \nu}=e_\mu^i e_\nu^j\nabla_{i} n_{j}$ representing the extrinsic curvature of the hyper-surface, and $\mathcal{K}=\mathcal{K}_{\mu}^{\mu}$ the corresponding trace of $\mathcal{K}_{\mu \nu}$ which will be used further down. Furthermore, using Eqs.\eqref{3a} to \eqref{5a} into the general expression for $\mathcal G$ one verifies that the Gauss-Bonnet invariant can be written as
\begin{align}\label{8a}&
\mathcal{G}=\mathcal{G}^{+} \mathbf{\Theta}(l)+\mathcal{G}^{-} \mathbf{\Theta}(-l)+\bar{\mathcal G}\delta\left(l\right),
\end{align}

Now we turn to the impact of the scalar fields $\Phi$ and $\Psi$. Thus these fields are written in the distribution formalism in the usual way as:
\begin{align}\label{16a}&
\Phi=\Phi^{+} \mathbf{\Theta}(l)+\Phi^{-} \mathbf{\Theta}(-l)
,
\\\label{17a}&
\Psi=\Psi^{+} \mathbf{\Theta}(l)+\Psi^{-} \mathbf{\Theta}(-l) .
\end{align}
Since the gravitational field equations mentioned in Eq.\eqref{18i} are based on the $2nd$-order derivatives of the scalar field $\Phi$, these terms must be examined in this formalism. When we take the partial derivative of Eq.\eqref{16a}, we get
\begin{align}\label{23a}
\partial_{i} \Phi=\Phi^{+} \mathbf{\Theta}(l)+\Phi^{-} \mathbf{\Theta}(-l)+\varepsilon[\Phi] n_{i} \delta(l).
\end{align}
Moreover, in this manuscript, we are concerned with an equivalent scalar-tensor representation of the $f(\mathcal{G},~T)$ gravity mentioned by the action in Eq.\eqref{17i}, instead of from a generic scalar-tensor theory provided by the action in Eq.\eqref{1i}. This interpretation is only characterized when the determinant of $\mathcal{A}$ in Eq.\eqref{matrixeq} does not vanish, as explained in the previous section. This definition states that both the scalar fields $\Phi$ as well as $\Psi$ can be expressed explicitly in the form of the Gauss-Bonnet invariant $\mathcal{G}$ and the trace of the stress-energy tensor $T$, i.e., $\Phi=\Phi(\mathcal{G},~T)$ and $\Psi=\Psi(\mathcal{G},~T)$, respectively, and vice versa, i.e., $\mathcal{G}=\mathcal{G}(\Phi,~\Psi)$ and $T=T(\Phi,~\Psi)$. With these arguments in mind, the $1st$, and $2nd$-order covariant derivatives of $\Phi$, can be written as:
\begin{align}\label{24a}&
\partial_{i} \Phi=\Phi_{\mathcal{G}} \partial_{i} \mathcal{G}+\Phi_{T} \partial_{i} T
,
\\\label{25a}
\nabla_{i} \nabla_{j} \Phi=\Phi_{\mathcal{G}} \nabla_{i} \nabla_{j} \mathcal{G}+\Phi_{T} \nabla_{i} \nabla_{j} T&+\Phi_{\mathcal{G} \mathcal{G}} \partial_{i} \mathcal{G} \partial_{j} \mathcal{G}+\Phi_{T T} \partial_{i} T \partial_{j} T+2 \Phi_{\mathcal{G} T} \partial_{(i} \mathcal{G} \partial_{j)} T,
\end{align}
where the subscripts $\mathcal{G}$ as well as $T$ stand for partial derivatives of these quantities, respectively. As a result, the existence of products of such manner $\partial_{i}\mathcal{G} \partial_{j} \mathcal{G}$, $\partial_{i}T \partial_{j} T$, and $\partial_{i} \mathcal{G} \partial_{j} T$ as in expression for $\nabla_{i} \nabla_{j} \Phi$ in Eq. \eqref{25a} suggests that even these differential terms would be dependent on products of the form $\mathbf{\Theta}(l) \delta(l)$ as well as $\delta^2(l)$, which are ill-defined and singular. To avoid these terms, the $\delta(l)$ terms in Eqs. \eqref{26a} and \eqref{27a} must be pushed to disappear, i.e., $[\mathcal{G}]=0$ and $[T]=0$. Furthermore, two JCs are implied by the conditions $[\mathcal{G}]=0$ and $[T]=0$, because both fields $\Phi$ and $\Psi$ are well-behaved functions of $\mathcal{G}$ and $T$ according to the definition of the equivalent scalar-tensor illustration even though discussed earlier in this paper.
\begin{align}\label{28a}
&{[\Phi]=0,} \\\label{29a}
&{[\Psi]=0.}
\end{align}
As a result, in Eq.\eqref{23a}, the $1st$-order derivative of $\Phi$ gets the form:
\begin{align}\label{30a}
\partial_{i} \Phi=\Phi^{+} \mathbf{\Theta}(l)+\Phi^{-} \mathbf{\Theta}(-l),
\end{align}
and we can now evaluate the $2nd$-order derivatives of $\Phi$, which are widely used as \begin{align}\label{31a}
\nabla_{i} \nabla_{j} \Phi=\nabla_{i} \nabla_{j} \Phi^{+} \mathbf{\Theta}(l)+\nabla_{i} \nabla_{j} \Phi^{-} \mathbf{\Theta}(-l)+\varepsilon n_{i}\left[\partial_{j} \Phi\right] \delta(l).
\end{align}

Now we deal with the theory's matter section. To work out the features of the thin-shell, let us express the stress-energy tensor $T _{i j}$ as a distribution function. The stress-energy tensor is represented as a distribution function of the form: $T_{(sr) \ ij}$ in the scalar-tensor representation, which we shorten as $T_{ij}$ to simplify notation i.e.,
\begin{align}\label{13a}
T_{i j}=T_{i j}^{+} \mathbf{\Theta}(l)+T_{i j}^{-} \mathbf{\Theta}(-l)+\delta(l) \mathcal{S}_{i j},
\end{align}

On the right hand side of Eq.\eqref{13a} $1st$ and $2nd$ terms (i.e., $T^{+}_{i j}$, $T^{-}_{i j}$) are the regular terms of stress-energy tensor in $\mathbf{M}^{+}$ and $\mathbf{M}^{-}$ regions, respectively in the scalar tensor representation. Also, $\mathcal{S}_{i j}$ denotes the thin-shell's $4$-dimensional stress-energy tensor in the same scalar tensor representation, which may be expressed as a $3$-dimensional tensor at separation hyper-surface $\Sigma$ as
\begin{align}\label{14a}
\mathcal{S}_{i j}=\mathcal{S}_{\mu \nu} e_{i}^{\mu} e_{j}^{\nu}.
\end{align}
In this manuscript, both the field equations in Eq.\eqref{18i} as well as the scalar field equation in Eq.\eqref{20i} also reliant explicitly in the trace of $T_{i j}$ i.e., $T=g^{i j} T_{i j}$. We can deduce from the trace of Eq.\eqref{13a} that $T$ equals
\begin{align}\label{15a}
T=T^{+} \mathbf{\Theta}(l)+T^{-} \mathbf{\Theta}(-l)+\delta(l) \mathcal{S},
\end{align}
where $\mathcal{S}$ is defined as $\mathcal{S}=\mathcal{S} ^{i}_{i}$.
Since $\Phi$ and $\Psi$ are characterized with no dependency in the $\delta(l)$ distribution. The potential function $\mathbb{V}(\Phi,~\Psi)$ is a function of $\Phi$ and $\Psi$, without any proportionality to $\delta(l)$, that guaranteed to be regular and does not feature any singular products of distribution functions of the form $\mathbf{\Theta}(l)\delta(l)$ or $\delta^2(l)$. Consequently, any partial derivative of $\mathbb{V}$ will be guaranteed to have the same regularity. The left-hand sides of Eqs.\eqref{19iG} and \eqref{19iT}, which is $\mathbb{V}_{\Phi}$ and $\mathbb{V}_{\Psi}$, respectively, would not be dependent on $\delta(l)$. As a direct consequence of Eqs.\eqref{19iG} and \eqref{19iT}, one can extrapolate that the Gauss-Bonnet invariant $\mathcal{G}$ as well as the trace of the stress-energy tensor $T$ are independent of $\delta(l)$. In this representation the equations of motion for the scalar fields force the condition $\bar{\mathcal G}=0$, to preserve the regularity of the MFEs, one immediately follow that the extrinsic curvature $\mathcal{K_{\mu\nu}}$ must be continuous across $\Sigma$. Thus the second and third JCs of the theory are derived from the explicit configurations of these two variables in Eqs.\eqref{8a} as well as \eqref{15a}, respectively as
\begin{align}\label{9a}
\left[\mathcal K_{\mu\nu}\right]=0.
\end{align}
and also,
\begin{align}\label{18a}
\mathcal{S}=0.
\end{align}
This new JC in Eq.\eqref{9a} an extremely restricted JC and GR doesn't have it or any of the previously extended versions of GR except in the situation of smooth matching, i.e., matching without thin-shells, and it is coupled with an arbitrary action dependence in $\mathcal G$. All the quantities $\bar R_{ijpq}$, $\bar R_{ij}$ and $\bar R$ disappear as a result of this condition, as do all the terms proportional to $\delta\left(l\right)$ in Eqs.\eqref{3a} to \eqref{5a}, and these  quantities become regular at $\Sigma$. As a result, we get the usual forms.
\begin{align}\label{3aaa}
R_{ijpq}&=R^{+}_{ijpq} \mathbf{\Theta}(l)+R^{-}_{ijpq} \mathbf{\Theta}(-l)
,
\\\label{10a}
R_{ij}&=R^{+}_{ij} \mathbf{\Theta}(l)+R^{-}_{ij} \mathbf{\Theta}(-l)
,
\\\label{11a}
R&=R^{+} \mathbf{\Theta}(l)+R^{-} \mathbf{\Theta}(-l)
,
\\\label{12a}
\mathcal{G}&=\mathcal{G}^{+} \mathbf{\Theta}(l)+\mathcal{G}^{-} \mathbf{\Theta}(-l).
\end{align}

These findings are analogous with those achieved in the geometric illustration of the theory, as they are inferred in Eqs.\eqref{rabcdreg} and \eqref{gbreg}. As a result, the trace of the stress-energy tensor $T$ is formed as
\begin{align}\label{20a}
&T=T^{+} \mathbf{\Theta}(l)+T^{-} \mathbf{\Theta}(-l).
\end{align}
In the distribution formalism, one gets $\partial_{i} \mathcal{G}$ and $\partial_{i} T$ from Eqs.\eqref{12a} and \eqref{20a}, respectively.
\begin{align}\label{26a}
\partial_{i} \mathcal{G}=\partial_{i} \mathcal{G}^{+} \mathbf{\Theta}(l)+\partial_{i} \mathcal{G}^{-} \mathbf{\Theta}(-l)+\varepsilon[\mathcal{G}] n_{i} \delta(l), \\\label{27a}
\partial_{i} T=\partial_{i} T^{+} \mathbf{\Theta}(l)+\partial_{i} T^{-} \mathbf{\Theta}(-l)+\varepsilon[T] n_{i} \delta(l).
\end{align}

Now we deduce the singular part of the field equations derive an explicit expression for the stress-energy tensor $\mathcal{S}_{\mu \nu}$ in the form of the geometrical quantities. To do just that, plug the expressions of the numerous significant quantities from Eqs.\eqref{1a}, \eqref{16a}, \eqref{17a}, \eqref{31a}, \eqref{13a}, \eqref{3aaa}, \eqref{10a}, \eqref{11a}, and \eqref{20a} into the field equations in Eq.\eqref{20i} with mandatary condition $\mathcal{K}_{\mu\nu}=0$, as well as project the finding into $\Sigma$ with $e_{\mu}^{i} e_{\nu}^{j}.$ The result is as follows
\begin{align}\label{32a}
\left(8 \pi+\Psi_{\Sigma}\right) \mathcal{S}_{\mu \nu}=4\varepsilon\left(e^i_\mu e^j_\nu R_{\ i\ j}^{p\ q}-R^{pq}\textbf{h}_{\mu\nu}\right)n_p\left[\partial_{q} \Phi\right]+4(e^{i}_{\mu}e^{j}_{\nu}R_{ij}-\frac{1}{2}R \mathbf{h}_{\mu \nu})\varepsilon n^{p}\left[\partial_{p} \Phi\right].
\end{align}
By introducing the trace of Eq.\eqref{32a}, and using established result $\mathcal{S}=0$, which provides a more convenient shape of a new JC as
$$\left[\partial_{p} \Phi\right]=0,$$
which could then be reinserted into Eq.\eqref{32a} to suggest the term on the right-hand side to disappear. Consequently, the right-hand side of Eq.\eqref{32a} vanishes identically, which implies that $\mathcal S_{\mu\nu}=0$. Thus thin-shells cannot emerge in the $f\left(\mathcal{G},T\right)$ gravity theory under this formalism, as shown by this conclusion. In this theory, all junctions between two distinct space-times at a given hyper-surface $\Sigma$ must be smooth, which is a rather restricted finding. We can summarize that the JCs composed for this equivalent scalar-tensor representation of the $f(\mathcal{G},~T)$ gravitational theory at $\Sigma$ is made up of five different equations as
\begin{equation}
\centering
\left.\begin{split}
{\left[\mathbf{h}_{\mu \nu}\right]}=0,\\
{[\mathcal{K_{\mu\nu}}] }=0,\\
{[\Phi] }=0,\\
{[\Psi] }=0,\\
{\left[\partial_{p} \Phi\right]}=0.\\
\end{split}~~\right\}
 \end{equation}
It is worthy to notice that these JCs may
directly composed from a very significant Eq.\eqref{58i} by using the introduction
of basic definitions like $\Phi=f_{\mathcal{G}}(\mathcal{G},~T)$ as well as $\Psi=f_{T}(\mathcal{G},~T)$, which may emphasize the equivalence between these two representations of this considered gravity.

\section{Conclusion}
In general relativity, one must ensure that the induced metric at $\Sigma$ as well as the extrinsic curvature are both continuous for the matching between the two space-times. Matching between the interior and exterior regions is still conceivable even if the extrinsic curvature is not continuous, but at $\Sigma$ the appearance of a thin-shell is required. In the recent study of astrophysics and cosmology, widely investigation is done from a thermodynamic standpoint on these thin-shells. Moreover, the entropy of the shell has also been estimated through a variety of scenarios, including electrically charged and spinning shells. Thus the agreement of the $1st$ and $2nd$-fundamental formulations at hyper-surface $\Sigma$ is the appropriate matching condition in GR. As a consequence, the certain perfectly matched hyper-surface in GR comprises that if and only if the jump of the $2nd$-fundamental form disappears, and thus the acceptable jumps in the energy-momentum tensor seem to be identical to Ricci (and Riemann) tensors non-vanishing discontinuities.

In this manuscript, we utilized the perfect fluid $f(\mathcal{G},~T)$ theory of gravitation, we took the universal formulation of action integral including modified GFEs. In our considered $f\left(\mathcal{G}, T\right)$ gravity has two extra scalar-degrees of freedom, representing two arbitrary dependencies of the function $f$ in the quantities $\mathcal G$ and $T$. We established a dynamically equivalent scalar-tensor form that is identically equivalent by introducing two scalar-fields $\Phi$ as well as $\Psi$, and also interaction potential $\mathbb{V}\left(\Phi,\Psi\right)$. To offer this analysis, we used the variance of modified EHA that is restated in a dynamically identical scalar-tensor formulation which changes under two auxiliary fields as $\alpha$ and $\beta$. It is supplied by the variables $\alpha=\mathcal{G}$ and $\beta=T$ into Eq.\eqref{13i}, consequently, by putting these discoveries one surly reconstruct Eq.\eqref{1i} which proved the two representations of our considered gravity are equivalent. Furthermore, we derived MFEs applying the modification to modified EHA given in dynamically equivalent scalar-tensor structure for these scalar fields, and interaction potential. We also established a more convenient form of the equations of motion by using matter Lagrangian as $\mathcal{L}_{\mathrm{m}}=-P$.
Of course, every gravitational theory must indeed have its own particular set of JCs, which should be derived from the full set of its GFEs. The GR with a cosmological constant term was the very first suggested modified theory of gravity, but also this theory has the identical JCs as the GR has too. The JCs have indeed been conceived in some other theories. In this manuscript, we took the general approach assuming the possibility of a thin-shell arising at $\Sigma$ between the two space-times $\mathbf{M}^{\pm}$. However, our results establish that, for the distribution formalism to be well-defined, thin-shells are not allowed to emerge in the general version of this theory.

We constructed a set of JCs in the scenario of $f(\mathcal{G},~T)$ gravitational theory by starting from the MFEs given in Eq.\eqref{9i}. We considered perfect fluid configurations to develop JCs that consist of only a singular part i.e., the terms are corresponding to $\delta(l)$ because the appearance of these terms in the derivatives of the metric is problematic. Thus in the distribution formalism to avoid the occurrence of these problematic products, we imposed the continuity of the metric $g_{ij}$. We confirmed that there is a complete absence of any term proportional to $\delta^2(l)$ in the Gauss-Bonnet invariant actually which identically vanished and its $1st$-order partial derivatives. It means that indicating terms are regular and do not appear in the singular field equation. We did, however, show that the $2nd$-order derivatives of the $\mathcal{G}$ have a singular term proportional to $[\partial_{j}\mathcal{G}]$. Such types of terms made an appearance in singular field equations, quite specifically in terms containing $2nd$-order derivatives of the function $f$. We verified that the stress-energy tensor $(T)$ has a term proportional to $\delta(l)$, which appeared in our established JCs. Similarly, we also verified that $T$ and it is $1st$-order partial derivatives did not have any singular terms and thus, not appeared in the singular field equations. However, we established that the $2nd$-order covariant derivatives of $T$ have a singular term proportional to $[\partial_{j} T]$, and this term appeared in the singular field equations, more precisely in the terms depending on $2nd$-order derivatives of $f$. Since in the MFEs, we have terms proportional to $2nd$-order derivatives of $f$, i.e., $\nabla_{i}\nabla_{j} f_\mathcal{G}$ and $\nabla^2 f_\mathcal{G}$. Thus, we showed that the terms proportional to $\nabla_{i}\nabla_{j} f_\mathcal{G}$ in the singular field equations will reduce to $\varepsilon n^{p}(f_{\mathcal{G}\mathcal{G}}[\partial_{p}\mathcal{G}]+f_{\mathcal{G}T}[\partial_{p} T])$ and a similar expression for $\nabla^2 f_\mathcal{G}$ but with the indices contracted.
Finally, we performed a projection to the hyper-surface $\Sigma$ with the projection vectors $e_\alpha^{i}$ and $e_\beta^{j}$ also in the end we redefined the indices back to $\mu$ and $\nu$.
Since $n_{i}$ is orthogonal to $e_\alpha^{i}$ by construction, the term proportional to $n_{i}$ vanished upon contraction, and the metric $g_{i j}$ contracted with the projection vectors and gave the induced metric. It is a well-known fact that MFEs of $f(\mathcal{G},~T)$ gravity show explicitly dependence in $T$ as well as its partial derivatives via the differential terms appear in $f_{\mathcal{G}}$ which refers to additional JCs.
Thus we used the trace of Eq.\eqref{58i} and established fact that $\mathcal{S}=0$ to get JC in form as $f_{\mathcal{G}\mathcal{G}}\left[\partial_{p} \mathcal{G}\right]+f_{\mathcal{G}T}\left[\partial_{p} T\right]=0.$
It means that the right-hand side of Eq.\eqref{58i} vanished identically, which implies that $\mathcal S_{\mu\nu}=0$. Because of this significant result, we found that thin-shells cannot emerge in the $f\left(\mathcal G, T\right)$ gravitational theory using this formalism. This is an extremely restricted conclusion in this theory, which requires that all junctions between two distinct space-times $\mathbf{M}^{+}$ and $\mathbf{M}^{-}$, at $\Sigma$ be smooth.

The same findings are achieved in the theory's scalar-tensor representation, demonstrating the equivalence of these two representations. This scalar-tensor depiction can only be characterized when the Hessian matrix of $f(\mathcal{G},~T)$ function is invertible, which coincides with the invertibility of the functions $\Phi(\mathcal{G},~T)$ and $\Psi(\mathcal{G},~T)$, the scalar fields and should be continuous. The partial derivatives of the scalar fields are forced to be continuous by the JCs for the trace $\mathcal{S}=0$, which implies that the $\mathcal{S}_{i j}$ dependency is not recovered entirely in the lump of metal as the extrinsic curvature $\mathcal{K}_{i j}=0$ in this formalism. Consequently, smooth matching proved the comparability of the two representations by recovering the continuity of the extrinsic curvature and the partial derivatives of the scalar field. In a dynamically equivalent scalar-tensor illustration, we have established a full set of JCs for smooth matching at $\Sigma$ under the same fluid version of $f(\mathcal{G},~T)$ gravitational theory.
Our findings significantly constrain the possibility of evolving models for alternative compact objects supported by thin-shells in this gravitational theory, such as gravastars and thin-shell wormholes, but they do provide a suitable framework for the search of models with a smooth matching at their surface, of which perfect fluid stars seem to be plausible examples. The results obtained indicate that under this formalism, thin-shells can not exist in the $f\left(\mathcal G,T\right)$ gravitational theory. This is a very restrictive result that forces all the junctions between two different space-times at $\Sigma$ to be smooth in this theory. There might be particular forms of the function $f\left(\mathcal G,T\right)$ for which some junction conditions can be discarded and maybe obtain thin-shells.

%%%%%%%%%%%%%%%%%%%%%%%%
%%%  Acknowledgments
%%%%%%%%%%%%%%%%%%%%%%%%
\section*{Acknowledgments}

We would like to thank Dr. Jo\~{a}o Lu\'{i}s Rosa from University of Tartu, Estonia for his
many valuable suggestions and comments that has significantly improved
the paper. 

\vspace{0.5cm}

\bibliography{mynewbibwt}
\bibliographystyle{ieeetr}
\end{document}